# Decarbonizing China's private passenger vehicles: A dynamic material flow assessment of metal demands and embodied emissions

Junhong Liu [1], Nan Zhou [2], Minda Ma [1 *], Kairui You [3 *]

1. School of Architecture and Urban Planning, Chongqing University, Chongqing, 400045, P. R. China
2. Energy and Resources Group, University of California, Berkeley, CA 94720, United States
3. The Center for Energy & Environmental Policy Research, Beijing Institute of Technology, 100081, Beijing, P.R. China

- Corresponding author: Prof. Dr. Minda Ma, Email: minda.ma@cqu.edu.cn
  Homepage: https://globe2060.org/MindaMa/

- Corresponding author: Dr. Kairui You, Email: youkairui@bit.edu.cn

## Graphical abstract

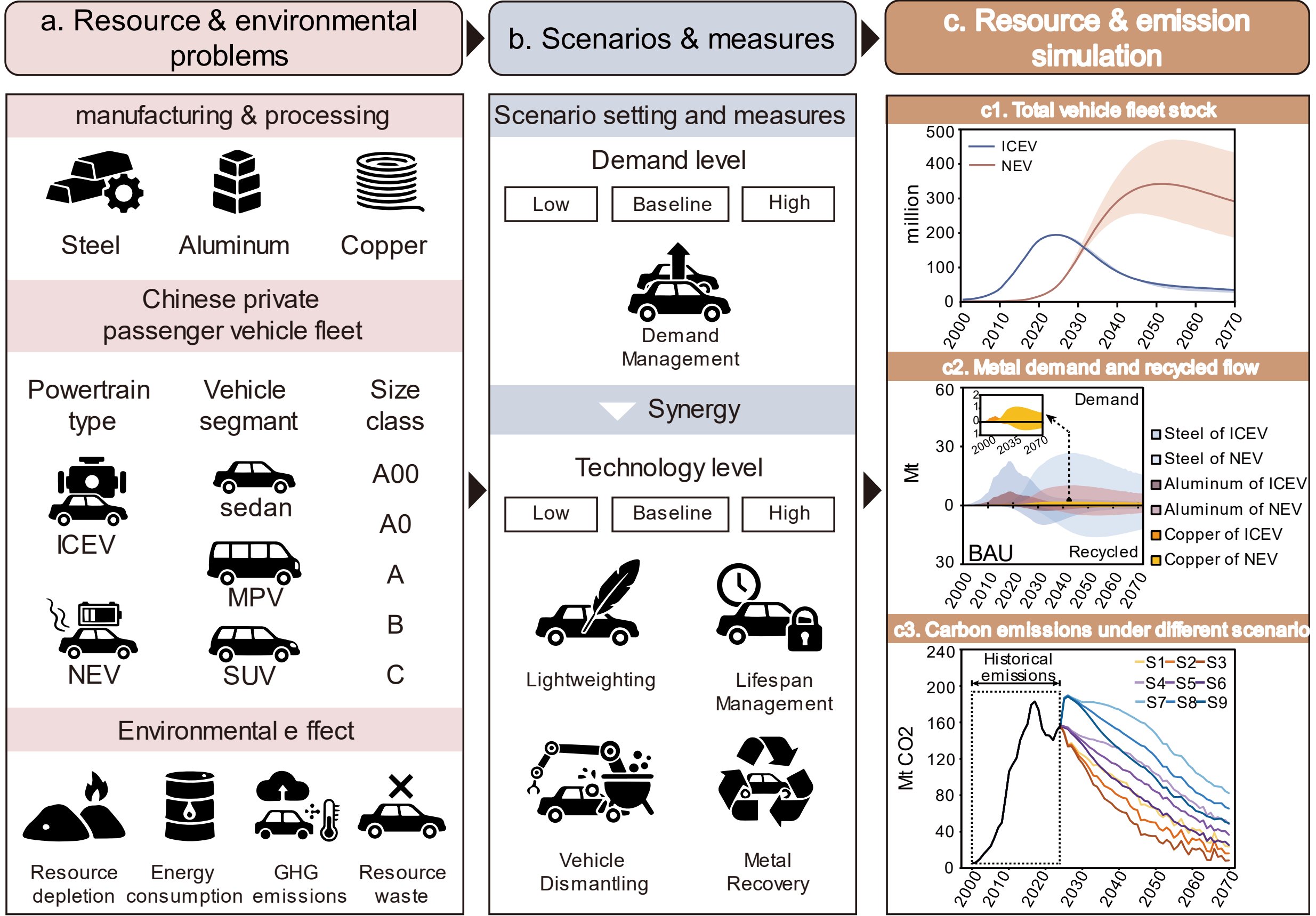

**Graphical Abstract.** A metal use efficiency modeling framework for China's private passenger vehicle fleet. The graphic illustrates: (a) the role of metals in the manufacturing and processing stages of passenger vehicles; (b) scenarios and measures designed to enhance metal use efficiency; (c1) total stock of China's private passenger vehicle fleet; (c2) metal demand and recycled metal flows; and (c3) embodied carbon emissions under different scenarios in 2000–2070.

## Highlights

- Developed a transferable dynamic MFA framework linking vehicle stocks, metal flows, and emissions.
- Passenger vehicle fleets peak at 327–507 million by mid-century, with NEVs dominating by the 2040s.
- Cumulative metal demand reaches 1914–2990 Mt by 2070, with up to 1320 Mt supplied from recycling.
- Embodied emissions reach 4958–9218 Mt $CO_2$ by 2070, depending on demand and technology levels.
- Technology upgrades cut 1051–1619 Mt $CO_2$, while demand management delivers 64.3% of reductions.

## Abstract

The continuous growth of China's private passenger vehicle fleet has intensified material demand and embodied carbon emissions, underscoring the need for effective decarbonization pathways. This study develops a transferable, dynamic material flow analysis framework to assess vehicle stocks, metal flows (steel, aluminum, and copper), and embodied emissions from 2000 to 2070, and to quantify the contributions of demand-side and technology-side efficiency measures. The results reveal that: (1) The vehicle fleet is projected to peak at 327–507 million vehicles by mid-century, with new energy vehicles dominating both in-use stocks and end-of-life flows by the 2040s. (2) Cumulative metal demand is projected to reach 1914–2990 million tonnes over the upcoming five decades, with 879–1320 million tonnes supplied from secondary sources under baseline conditions. Technology-oriented measures substantially enhance recycling performance, enabling secondary steel to fully meet manufacturing demand and allowing aluminum and copper cycles to approach near closure by 2070. (3) Correspondingly, cumulative embodied carbon emissions from vehicle metals by 2070 range from 4958 to 9218 megatonnes of carbon dioxide, with technological upgrading reducing emissions by 1051–1619 megatonnes. In collaborative scenarios, demand management accounts for 64.3% of total emission reductions, while technology-oriented measures become increasingly important over the medium to long term. Overall, the findings demonstrate that unmanaged demand growth can substantially offset technological mitigation gains, highlighting the necessity of integrated demand- and technology-oriented strategies. This study provides a systemic and transferable framework to guide circular economy development and deep decarbonization transitions in vehicle fleets in China and other emerging economies.

## Keywords

Private passenger vehicle; Embodied carbon emissions; Demand management; Metal use efficiency; Fleet electrification and lightweighting; Material flow analysis

### *Abbreviation notation*

BAU - Business-as-usual

$CO_2$ – Carbon dioxide

DMFA - Dynamic material flow analysis

EF - Emission factor

ELV - End-of-life vehicle

EOL - End-of-life

ICEV - Internal combustion engine vehicle

MI - Metal intensity

Mt - Megatonne

NEV - New energy vehicle

RIR - Recycling input rate

### *Nomenclature*

### Parameters and variables

$a$ - Saturation level of passenger vehicle ownership per 1000 people

$b$ - Scaling parameter controlling the vertical span and initial offset of logistic growth model

$c$ - Shape parameter determining the steepness of logistic growth model

$D_m(t)$ - Total demand for metal $m$ in year $t$

$E_{avoided}(t)$ - Emissions avoided from secondary metal substitution in year $t$

$E_{embodied}(t)$ - Embodied carbon emissions of metal $m$ in year $t$

$E_m(t)$ - Theoretical end-of-life availability of metal $m$ in year $t$

$E_{recycling}(t)$ - Emissions generated from the metal recycling process of ELVs in year $t$

$G_v(t)$ - Annual scrappage of vehicle category $v$ in year $t$

$k$ - Shape parameter of the Weibull distribution

$m$ - Metal type (steel, aluminum, or copper)

$M_{v,m}$ - Mass of metal $m$ contained in a single vehicle of category $v$

$R_m(t)$ - Amount of recycled metal $m$ available for reuse in year $t$

$Sales_{t,x}$ - Remaining in-use stock in year $t$ of vehicles sold in year $x$

$Sales_x$ - Total vehicle sales in year $x$

$Scrappage_{t,x}$ - Number of vehicles sold in year $x$ that are scrapped in year $t$

$Scrappage_t$ - Total vehicle scrappage in year $t$

$Stock_t$ - Total in-use vehicle stock in year $t$

$S_v(t)$ - Annual sales of vehicle category $v$ in year $t$

$t$ - Year observed in the model

$v$ - Vehicle category defined by powertrain type, vehicle segment, and size class

$x$ - Year in which vehicles are sold

$X$ – Independent variable, typically time (years) or a related metric

$Y$ – Vehicle ownership per 1000 people

**Greek letters**

$\alpha(t)$ - Vehicle dismantling rate in year $t$

$\Delta C(t)$ - Net carbon emission reduction achieved in year $t$

$\lambda$ - Scale parameter of the Weibull distribution

$\rho_m$ - Metal recovery efficiency for metal $m$

$\tau$ - Time-scale parameter of logistic growth model, indicating the peak growth year

**Emission factors**

$EF_{primary,m}$ - Emission factor for primary production of metal $m$

$EF_{secondary,m}$ - Emission factor for secondary production of metal $m$

## 1. Introduction

Over the past three decades, transport-sector emissions have risen sharply, now accounting for more than one-fifth of global carbon dioxide ($CO_2$) emissions [1], with passenger transport contributing about 60% due to its heavy reliance on internal combustion engine vehicles (ICEVs) [2]. As global passenger demand continues to grow [3], decarbonizing this sector is critical for achieving carbon neutrality. Although the transition from ICEVs to new energy vehicles (NEVs) can reduce operational emissions by up to 73% [4], the sector still faces significant energy and resource challenges. Rapid vehicle growth increases manufacturing-related energy use and emissions, driven by demand for energy-intensive metals such as steel, aluminum, and copper [5, 6], which together account for nearly two-fifths of the global carbon footprint of durable goods [7]. Moreover, rising vehicle ownership accelerates vehicle scrappage, leading to resource losses, waste generation, and environmental pollution [8], underscoring the need for more efficient resource management in vehicle production and recycling [9].

Since 2009, China has been the world's largest producer and market for passenger vehicles and has undergone an electrification transition to promote cleaner transport [10]. Although electrification reduces operational emissions [11], the full life cycle—including manufacturing and end-of-life (EOL) stages—continues to pose significant environmental challenges due to embodied carbon emissions [12]. Achieving substantial carbon reductions therefore requires not only cleaner energy sources but also improvements in metal use efficiency to reduce energy consumption and emissions across the entire life cycle. **However**, NEVs differ from ICEVs in terms of life-cycle characteristics, metal intensity (MI), and vehicle composition [13]. In addition, the shorter lifespan of NEVs accelerates metal turnover, thereby affecting metal demand. Combined with ongoing lightweighting trends, electrification is reshaping the flows of major metals such as steel, aluminum, and copper, underscoring the need for a holistic, system-level approach to carbon emission reduction. This study focuses on China's private passenger vehicle fleet and examines the demand for steel, aluminum, and copper, their recycling potential, and the carbon reduction benefits associated with improved metal use efficiency. Specifically, this study addresses the following research questions:

- How will China's private passenger vehicle fleet transition over the next five decades?
- How will the stock and flow of steel, aluminum, and copper within the vehicle fleet evolve?
- How can improvements in metal use efficiency reduce emissions across the vehicle life cycle?

To address these questions, this study develops a data-driven, dynamic model of China's private passenger vehicle fleet, incorporating cohort-level characteristics such as vehicle ownership, powertrain type, vehicle segment, and size class, as well as interactions among metals, energy, and emissions. The model simulates the stock and flow dynamics of steel, aluminum, and copper from 2025 to 2070. In addition, multi-scenario analyses are conducted to evaluate the impacts of technology-oriented and demand-oriented metal use efficiency measures on resource consumption and carbon emissions, thereby providing a scientific basis for low-carbon industrial transformation and evidence-based policymaking.

**The principal contribution of this study** lies in the development of a transferable, dynamic, system-level vehicle fleet metal flow model that integrates fleet electrification and lightweighting. This framework captures how emerging technologies reshape metal demand and life-cycle carbon emissions—an aspect that has been largely overlooked in previous studies. At the intervention level, the model simultaneously considers demand-oriented and technology-oriented metal use efficiency measures and quantifies their individual and combined contributions to carbon reduction. By explicitly linking technological transitions with actionable mitigation measures, the framework provides an enhanced and transferable tool for optimizing circular and low-carbon pathways in private passenger vehicle fleets, thereby bridging the gap between theoretical modeling and practical decarbonization planning.

The remainder of this paper is organized as follows. Section 2 reviews the relevant literature. Section 3 describes the methods and materials, including passenger vehicle fleet projections and carbon emission assessment. Section 4 presents projected transitions in passenger vehicle stock, metal flows, and carbon emissions under different scenarios. Section 5 discusses the carbon reduction potential of metal use efficiency measures and the associated policy implications. Section 6 summarizes the key findings and outlines directions for future research.

## 2. Literature review

Existing studies have consistently demonstrated that vehicle-related carbon emissions are largely determined in upstream stages rather than being solely concentrated in operational emissions during the use phase. The production of major metals such as steel and aluminum used in vehicle bodies contributes approximately 68% of the total energy consumption [14] and approximately 18–22% of the carbon emissions throughout the vehicle's entire life cycle [15]. Consequently, the embodied carbon emissions in the manufacturing phase should not be overlooked [16]. Within this context, two major technological transitions—lightweighting [17] and electrification [18]—have been widely discussed as key strategies for reducing vehicle-related emissions. Notably, while lightweighting and electrification effectively reduce operational emissions from vehicles [19], they may increase embodied carbon emissions.

A substantial body of literature has examined the emission implications of vehicle lightweighting. Vehicle lightweighting primarily reduces operational energy consumption and thus carbon emissions by lowering the vehicle weight [20]. Lightweighting designs typically replace traditional steel with materials such as aluminum, high-strength steel, and carbon fibers [21]. However, some researchers have noted that the production of these materials often results in a greater carbon footprint [22]. For example, China's primary aluminum production has a emission intensity of 16.0 tonnes of carbon dioxide equivalent (t$CO_2$-eq) per tonne, which is 6.4 times greater than that of steel. Therefore, the net environmental benefit of lightweighting depends on a trade-off between reduced operational emissions and increased manufacturing emissions, as emphasized in several life cycle assessment studies [23, 24].

Similar trade-offs are observed in studies on vehicle electrification. From 2005 to 2030, electrification is predicted to reduce greenhouse gas emissions from light-duty fleets by approximately 25% [25]; however, electrification also increases the demand for copper in motors and high-voltage wiring. The emission intensity of primary copper is 3.5–4.5 t $CO_2$-eq/t [26], and since electric vehicles use approximately 3–4 times more copper than conventional vehicles do [27], this leads to higher embodied carbon emissions during manufacturing. These findings indicate that electrification reshapes both energy use patterns [28] and upstream material demand structures.

Methodological advances have played a central role in quantifying such material-emission interactions (see Table 1). Material flow analysis is widely applied to quantify the stock and flow of major metals or minerals [29] and can be divided into static material flow analysis and dynamic material flow analysis (DMFA) [30]. Increasingly, scholars have combined DMFAs with life cycle assessments to quantify the life cycle emissions of major metals in China's passenger vehicle fleet [8, 31] and to assess the potential for secondary supply [32, 33]. While these studies have improved the understanding of past and current material flows, the DMFA is still largely retrospective or trend extrapolative [34].

Building on these methodological foundations, recent studies have increasingly examined potential future demand pathways [35, 36] and projected emission levels [37] for major metals, such as steel [38], aluminum [32], and copper [39], in China's passenger vehicle fleet by adopting different demand levels, technological development paths, and policy intensities. These scenario-based approaches extend the conventional DMFA from descriptive analysis to strategic planning and policy evaluation [40]. In addition to methodological improvements, improving resource use efficiency is another important focus. Achieving deep decarbonization requires balancing carbon emissions between the vehicle use and manufacturing stages [41]. Studies have shown that through a combination of measures such as extending vehicle lifespan, enhancing standardization and modularization, optimizing material substitution, and promoting closed-loop recycling [42, 43], resource consumption and emissions per unit of transport service function can be minimized [44]. At the fleet level, systematic resource management can decouple resource consumption and carbon emissions at the fleet level [45].

**Table 1.** Comparative analysis of existing DMFA studies on vehicle materials.

| Study | Methodological framework | Temporal scope | System boundary | Key features |
|---|---|---|---|---|
| [32] | Material flow analysis of aluminum alloys | 2019 to 2050 | Automotive aluminum production and recycling | Aluminum demand and recycling potential under electrification and lightweighting measures |
| [36] | Bottom-up battery material flow model with vehicle-segment heterogeneity | 2020 to 2050 | Electric vehicle batteries and critical materials | Battery material demand and recycling potential across vehicle segments |
| [46] | DMFA integrated with national energy technology model | 2020 to 2060 | Electric vehicle batteries and critical metals supply chain | Critical metal demand and supply risks for EV batteries under circular economy strategies |
| [47] | DMFA combined with LMDI decomposition and Gaussian process regression forecasting | 2023 to 2030 | Provincial NEV battery metal demand in China | Regional heterogeneity and driving factors of critical metal demand |
| [48] | DMFA combined with lifespan distribution modeling and Markov chain analysis | 2015 to 2050 | U.S. automotive aluminum body sheet flows from aluminum-intensive vehicles | Future generation, composition, and recycling potential of aluminum body sheet scrap |

**However, despite the valuable insights provided by existing studies into the demand for critical metals during the transition to vehicle electrification, several notable gaps remain.** First, the research focus is uneven: most studies emphasized critical metals associated with EV batteries, such as lithium, cobalt, and nickel, while comparatively little attention has been paid to structural metals in vehicle bodies, such as steel and aluminum, and their transformation under electrification. Second, the temporal scope of prior work is generally limited (typically covering 2020–2050), constraining the ability to fully capture the long-term dynamic effects of the large-scale replacement of ICEVs with NEVs, including technological generational shifts and subsequent EOL recycling waves that influence metal supply and demand. Finally, only a limited number of studies have integrated the two core transition pathways—vehicle lightweighting and fleet electrification—within a unified framework,

systematically examining their coupled effects and the resulting implications for the full life cycle of metals.

**To address these gaps, this study develops a bottom-up, dynamic vehicle fleet model that captures the combined effects of vehicle lightweighting, fleet electrification, and improvements in metal use efficiency.** The model incorporates emission factors (EFs) for both primary and secondary metals to quantify manufacturing-stage carbon emissions over the period 2025–2070. It further evaluates the mitigation potential of key metal use efficiency strategies—including lightweighting, vehicle lifespan extension, and enhanced end-of-life vehicle (ELV) dismantling—in advancing embodied carbon neutrality. The specific contributions of this work are summarized as follows:

- **This study is the first to develop a long-term, dynamic, system-level model** that explicitly simulates the synergistic effects of vehicle lightweighting and fleet electrification on multi-metal demand and embodied carbon emissions. The framework addresses key limitations of existing studies in terms of temporal coverage and system boundaries, while incorporating high parameter granularity and scenario-based quantification. This enables explicit modeling of substitution effects and circular economy contributions across multiple vehicle segments and metal types.
- **This study provides practical insights for low-carbon transition strategies in the transport and materials sectors.** By addressing gaps in long-term, multi-process integration and carbon-neutral pathway assessment, the framework supports evidence-based metal resource management and informs decarbonization strategies for China's private passenger vehicle fleet, with clear applicability to other emerging economies.

## 3. Methods and Materials

This study aims to develop an integrated methodological framework to evaluate the decarbonization potential of China's private passenger vehicle fleet from a metal cycling perspective (see Fig. 1). Section 3.1 establishes a passenger vehicle fleet projection model that simulates fleet growth, sales, and scrappage. Section 3.2 presents a DMFA model to quantify major metal flows along demand- and technology-oriented transition pathways. Section 3.3 describes the carbon emission assessment model, which integrates EFs for primary and secondary metals with DMFA results. Section 3.4 outlines the scenario and measurement design, and Section 3.5 summarizes the main data sources.

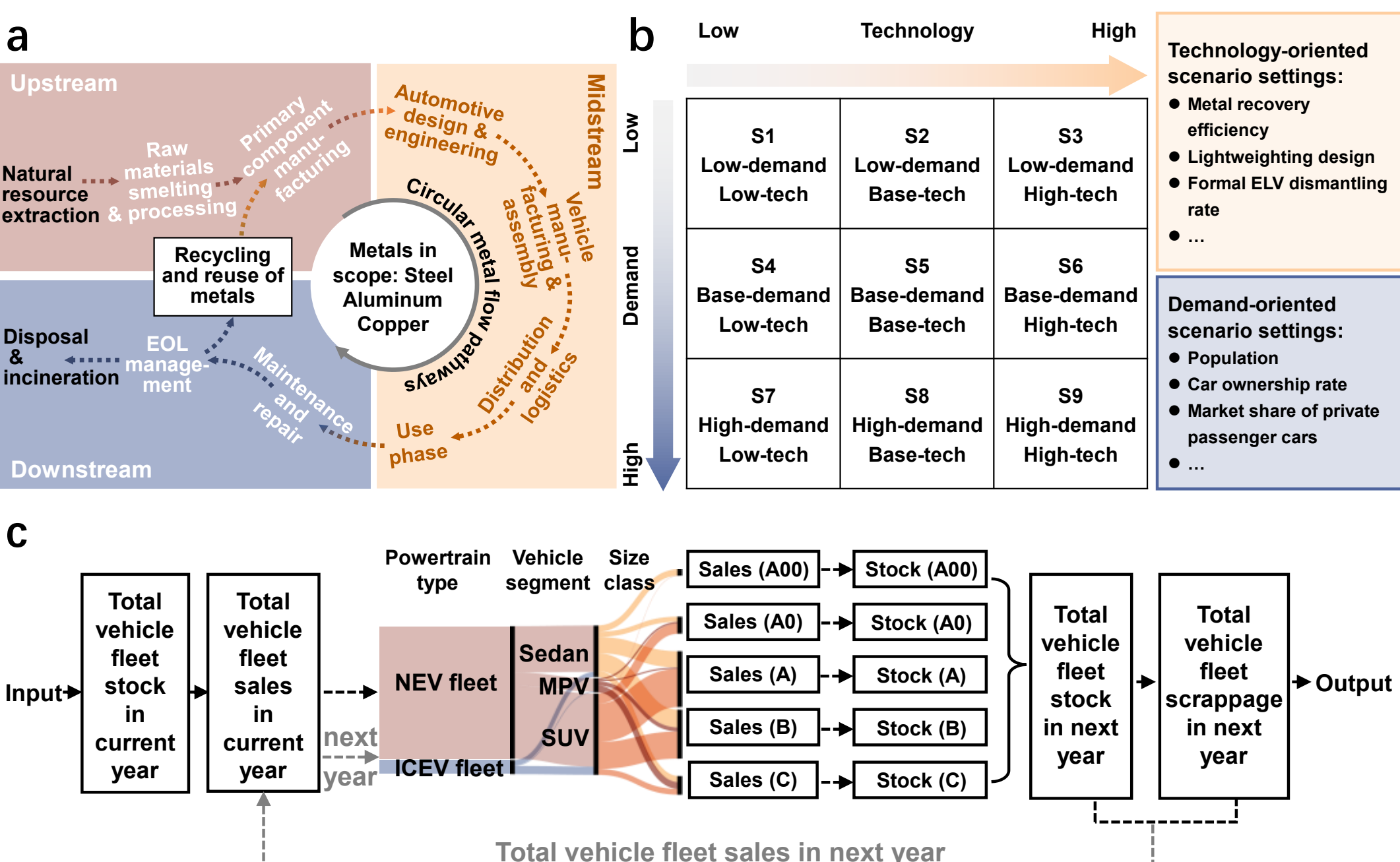

**Fig. 1.** Overview of the research framework: (a) metal life-cycle processes within the private passenger vehicle fleet; (b) scenario design and parameterization of vehicle fleet transition pathways; and (c) computational workflow for projecting passenger vehicle fleet stocks.

### *3.1. Projection of passenger vehicle fleet demand*

This study used a logistic growth model to predict the long-term saturation trajectory of China's private passenger vehicle fleet. This S-shaped diffusion model is suitable for characterizing the popularity dynamics of durable goods approaching market saturation. The basic function of the logistic model is expressed as follows:

$$Y = a - \frac{b}{1 + (X/\tau)^c} \tag{1}$$

Here, $a$ represents the upper asymptote, indicating the ultimate saturation level of vehicle ownership per 1000 people. $b$ is a scaling parameter controlling the vertical span and initial offset of the curve. $\tau$ denotes the time-scale parameter, which is closely associated with the inflection point of the S-curve where growth is fastest. $c$ is the shape parameter, which was used to determine how abruptly the growth rate transitions toward saturation. In reference to the level of passenger vehicle ownership per 1000 people in developed countries (such as over 800 vehicles per 1000 people in the United States, approximately 642 in Japan, 501 in South Korea, and 371 in Russia [49]) and considering China's development stage and policy context (approximately 210 vehicles per 1000 people at present), this study set 400 vehicles per 1000 people as the saturation benchmark, representing a moderate growth trajectory consistent with China's anticipated development pathway and policies aimed at limiting private vehicle ownership. The low-demand scenario assumes a more conservative growth path, with a saturation level of 350 vehicles per 1000 people, reflecting intensified urbanization, stricter policy constraints, and greater investment in public transportation. In contrast, the high-demand scenario adopts a saturation level of 450 vehicles per 1000 people, representing sustained growth in private vehicle ownership driven by urban sprawl and continued policy support. By fitting historical data from 2000–2025 through nonlinear regression, this study projected annual per capita vehicle ownership until 2070. In addition, China's future population projections, derived from the low-, medium-, and high-variant scenarios in the *United Nations World Population Prospects 2024*, were used to estimate the total size of the private passenger vehicle fleet.

The fleet projection model in this study was built upon historical national total vehicle stock data and employed an iterative approach to simulate dynamically the relationships among stock, sales, and scrappage volumes. Specifically, this study assumed that the vehicle lifespan follows a Weibull distribution. The scrappage volume $Scrappage_{t,x}$, defined as the number of vehicles sold in year $x$ that are scrapped in year $t$, and the remaining stock $Sales_{t,x}$, defined as the number of vehicles sold in year $x$ that are still in use in year $t$, are formulated as follows:

$$Scrappage_{t,x} = Sales_x \cdot \left[\exp\left(-\left(\frac{t-x-1}{\lambda}\right)^k\right) - \exp\left(-\left(\frac{t-x}{\lambda}\right)^k\right)\right] \tag{2}$$

$$Sales_{t,x} = Sales_x \cdot exp\left[-\left(\frac{t-x}{\lambda}\right)^k\right] \tag{3}$$

where $\lambda$ and $k$ represent the scale parameter and shape parameter of the Weibull distribution, respectively. For vehicles, $k$ is typically set to 2.5, which is commonly used in vehicle lifespan modeling to reflect the distribution of vehicle age at the time of scrappage.

To estimate the total vehicle stock $Stock_t$ in year $t$ from all historical sales, it is necessary to sum the sales data from all previous years via the formulation above:

$$Stock_t = \sum_{x=t_{min}}^{t} Sales_x \cdot exp\left[-\left(\frac{t-x}{\lambda}\right)^k\right] \tag{4}$$

The sales volume $Sales_t$ in year $t$ is calculated as follows:

$$Sales_t = Stock_t - Stock_{t-1} + Scrappage_{t-1} \tag{5}$$

Building upon this framework, the annual total sales volume was integrated with projected NEV penetration rates, which were derived from policy targets, historical trends, and cost forecasts, to allocate new vehicle sales between ICEVs and NEVs. These sales figures were further disaggregated by vehicle type, with adjustments informed by historical market shares and evolving consumer preferences. This granular approach provides a robust foundation for subsequent modeling of material intensity (MI) and embodied carbon emissions.

### *3.2. DMFA model*

The core objective of the DMFA model is to translate the flow of vehicle stocks—predicted by the fleet projection model—into corresponding flows of major metals, thereby quantifying the future demand for primary metals and the availability of secondary metals (metals recycled from ELVs and reintroduced into the production system). Each vehicle was categorized along three dimensions: powertrain type (ICEV or NEV), vehicle segment (sedan, sport utility vehicle (SUV), or multi-purpose vehicle (MPV)), and size class (A00, A0, A, B, C), creating unique vehicle categories denoted as $v$. Market share distribution across vehicle categories is provided in Appendix B.

A critical component of the DMFA model is the MI matrix, where each parameter $Mv,m$ (units: kilogram/vehicle) represents the embedded mass of metal $m$ (steel, aluminum, or copper) in a single vehicle of category $v$. The detailed parameter values defining the MIs for each vehicle category are provided in Appendix C. These MI values were derived from the technical literature, industry benchmarking reports, and vehicle teardown studies and are further adjusted to incorporate projected lightweighting trends, thereby reflecting evolving metal use patterns over time.

For annual primary metal demand calculations, the model processes outputs from the fleet turnover model. The total demand for metal $m$ in year $t$, denoted $D_m(t)$, was calculated by aggregating the metal requirements across all vehicle categories. Specifically, the annual sales $S_v(t)$ of each vehicle category $v$ were multiplied by its corresponding MI $M_{v,m}$ and summed over all categories according to the following formula:

$$D_m(t) = \sum_{v} S_v(t) \cdot M_{v,m} \tag{6}$$

Similarly, the annual EOL metal availability for metal $m$ was estimated on the basis of scrappage flows. The EOL metal flow $E_m(t)$ represents the theoretical quantity of metal $m$ embedded in scrapped vehicles in year $t$. It was calculated by multiplying the annual scrappage $G_v(t)$ of each vehicle category $v$ by its corresponding MI $Mv,m$ and summing across all categories:

$$E_m(t) = \sum_{v} G_v(t) \cdot M_{v,m} \tag{7}$$

To estimate the actual supply of secondary metals, the theoretical EOL availability $E_m(t)$ was further adjusted to account for losses occurring during vehicle collection and recycling processes. In practice, not all deregulated vehicles enter the formal dismantling and recycling system, and not all metals contained in dismantled vehicles are successfully recovered. To capture these effects, the model applied a two-step correction: (1) The dismantling rate $\alpha(t)$ is defined as the fraction of officially scrapped vehicles that are collected and processed within the formal recycling chain in year $t$. (2) Metal recovery rate $\rho_m$: represents the mass recovery efficiency of the metal $m$ during recycling. This parameter reflects differences in recycling technologies and process efficiencies among metals. Accordingly, the amount of secondary metal actually available for reuse for each metal $m$ in year $t$, denoted as $R_m(t)$, is therefore calculated as:

$$R_m(t) = E_m(t) \cdot \alpha(t) \cdot \rho_m \quad (8)$$

These calculations were performed annually for each year from 2000–2070 and for each metal $m$, generating comprehensive annual time series of primary metal demand and secondary metal supply. This bottom-up DMFA approach ensures sensitivity to changes in vehicle fleets, including changes across powertrain types, vehicle segments, and size classes, and provides essential input for subsequent assessments of environmental impacts, resource security, and circular economy strategies, particularly with respect to identifying future secondary metal potentials and reducing primary metal requirements.

### *3.3. Embodied carbon emission assessment*

To quantitatively assess the carbon mitigation potential of different metal use efficiency measures, a carbon emission assessment model was developed. The core principle of this model is to compare the life cycle carbon emissions associated with metal production for vehicle manufacturing under two systems: (1) a baseline scenario reliant on primary metal production and traditional manufacturing and (2) various intervention scenarios incorporating secondary metal substitution through metal use efficiency measures.

The net carbon emission reduction achieved by an intervention measure is defined as the difference between avoided emissions from primary metal production and the emissions generated during metal recycling processes. Accordingly, the annual net carbon emission reduction is defined as:

$$\Delta C(t) = E_{avoided}(t) - E_{recycling}(t) \quad (9)$$

where $\Delta C(t)$ represents the net carbon emission reduction in year $t$ (kilogram of $CO_2$). $E_{avoided}(t)$ represents emissions avoided through the substitution of primary metals with secondary metals in year $t$, which constitute the dominant source of carbon savings in the vehicle manufacturing stage. $E_{recycling}(t)$ represents the emissions generated during the metal processing stages of ELV recycling, including dismantling, shredding, sorting, and remelting, in year $t$.

For any given measure or combination of measures, the avoided emissions are calculated by integrating the secondary metal supply obtained from the DMFA model with metal-specific EFs. In this study, the emission factors for primary and secondary metals were dynamically adjusted to reflect the projected decarbonization of the electricity grid, thereby capturing the anticipated decline in China's grid carbon intensity over time. The avoided emissions $E_{avoided}(t)$ associated with secondary metal substitution are expressed as:

$$E_{avoided}(t) = \sum_{m} R_m(t) \cdot (EF_{primary,m} - EF_{secondary,m}) \tag{10}$$

where $R_m(t)$ is the amount of recycled metal $m$ available for use in year $t$.$EF_{primary,m}$ represents the EF for producing metal $m$ from primary ore via the long-process route, including mining, concentration, smelting, and refining. $EF_{secondary,m}$ denotes the EF for producing metal $m$ from secondary sources via the short-process route, such as remelting and refining. The specific EF values for each metal can be found in Appendix D.

The emissions generated during recycling are calculated on the basis of the actual amount of metal entering the formal recycling system, which is consistent with the secondary metal supply defined in the DMFA model. The recycling emissions are therefore expressed as:

$$E_{recycling}(t) = \sum_{m} E_m(t) \cdot \alpha(t) \cdot EF_{secondary,m} \tag{11}$$

where $E_m(t)$ represents the theoretical EOL metal $m$ availability embedded in scrapped vehicles in year $t$ and where $EF_{recycling,m}$ denotes the EF associated with recycling metal $m$, encompassing emissions from dismantling and metal processing operations.

In addition, the embodied carbon emissions of the metal $m$ in year $t$, $E_{embodied}(t)$, can be calculated by multiplying the primary metal demand with its primary EF. The formula for the embodied carbon emissions is as follows:

$$E_{embodied}(t) = \sum_{m} D_m(t) \cdot EF_{primary,m} \tag{12}$$

This formulation ensures internal consistency between metal flow accounting and carbon emission assessment, avoids double counting, and explicitly links secondary metal availability to both avoided and recycling-related emissions. Overall, this model provides a robust and transparent framework for comparing the relative effectiveness of alternative EOL management and circular economy strategies in decarbonizing the automotive metal cycle.

### *3.4. Scenario design for metal use efficiency*

To evaluate the future carbon reduction pathways of China's passenger vehicle industry, this study developed a set of scenario combinations that include demand-side and technology-side measures and further quantified the independent contribution of each individual measure through individual factor analysis. All analyses were conducted by adjusting the core input parameters corresponding to the vehicle fleet projection model, DMFA model, and carbon emission models mentioned in Sections 3.1–3.3. Scenario integration aims to explore the system-level emission reduction potential under the synergistic effect of demand-side and technology-side measures. Each scenario was defined by a combination of two parameter dimensions, demand level (low, baseline, high) and technology level (low, baseline, high), resulting in nine scenario combinations that simulate diverse future trajectories under different levels of action across these two key levers. When both demand and technology are at the baseline level, it represents the business-as-usual (BAU) scenario.

The demand side is affected by one measure, namely, demand management, which was characterized by the number of passenger vehicles per 1000 people. The low demand level sets a lower ownership target, representing a future pathway of intensive urban development and strong transportation demand management. The baseline demand level adopts a benchmark ownership trajectory on the basis of the extrapolation of current trends, representing a conventional pathway that continues existing policies and development models and serves as the benchmark scenario for this study. The high demand level sets a higher ownership level, representing a pathway of sustained growth in the dependence on private passenger vehicles. The specific parameter settings for the levels are shown in Appendix E.

The technical side was determined by four key measures: lightweighting level, average service life, vehicle dismantling and recycling rate, and metal recovery efficiency. The low technology level assumes that all four parameters mentioned above are set to relatively lower values, reflecting delays in technological progress and policy implementation. The baseline technology level sets these four parameters at the benchmark level, which is consistent with current industry planning. Under the high-technology level, these four parameters are set to relatively high values, representing an advanced scenario in which technological breakthroughs are achieved and the circular economy system is highly developed. The parameter settings for the technology-side measures are shown in Appendix F.

### *3.5. Data collection*

The data used in this study were obtained from multiple authoritative sources to ensure accuracy and comprehensiveness. Historical annual vehicle ownership statistics for China were sourced from official publications of the Ministry of Public Security of the People's Republic of China [50]. National population projections were retrieved from the United Nations *World Population Prospects 2024* [51]. The market share of private passenger vehicles was calculated on the basis of data presented in the *China Automobile Market Outlook 2025*. Data regarding the minimum stock of ICEVs as well as sales figures for NEVs were acquired from the China Passenger Car Association, which was accessible via the official website (http://www.cpcaauto.com). Lifespan parameters for both ICEVs and NEVs were derived from experimental data provided by the China Automotive Technology and Research Center, with certain parameters supplemented by author estimates where empirical data were limited. Furthermore, MI metrics for various vehicle models were compiled from publicly available specifications on the automotive information platforms Chezhuzhijia (www.autohome.com.cn) and Dongchedi (www.dongchedi.com), whereas values for less documented models were estimated on the basis of technological analogs and engineering specifications.

## 4. Results

### *4.1. Long-term dynamics of private passenger vehicle stock, sales, and scrappage*

Fig. 2 indicates that the total number of private passenger vehicles in China will continue to grow until the middle of this century. Under the BAU scenario, total stock was projected to increase from approximately 245 million vehicles in 2025 to a peak of 393 million vehicles in 2047, followed by a gradual decline. Scenario analysis further reveals that demand intensity strongly affects the overall size of China's private passenger vehicle market, with peak total stock ranging from 327 to 507 million vehicles across scenarios (Fig. 2a). ICEVs and NEVs play fundamentally different roles in shaping these dynamics. The ICEV market has entered a mature and declining stage, with stock, sales, and scrappage exhibiting stable and predictable downward trajectories across all scenarios. By 2070, ICEV stock was expected to stabilize within a narrow range of 24.5–32.7 million vehicles. In contrast, the NEV market represents the primary source of future uncertainty and growth potential. Its development pathway directly determines the peak level of China's total private passenger vehicle stock, resulting in a wide projected range of 184.6–433.4 million vehicles by 2070.

Based on historical sales data, ICEV sales peaked at 24.5 million vehicles in 2017. By 2025, annual NEV sales were projected to surpass ICEV sales and continue increasing, reaching a peak of approximately 34.1 million vehicles around 2043. Moreover, pronounced differences in product lifespans between NEVs and ICEVs, together with accelerated shifts in fleet composition, not only shape the overall scale of future scrappage waves but also influence their timing. Model results indicate that total scrappage of China's private passenger vehicle fleet will reach a structural peak around 2046, at approximately 37.0 million vehicles per year. Of this total, NEVs were projected to account for about 84% (approximately 29.8 million vehicles), while ICEVs contribute only 16% (approximately 7.2 million vehicles). This pattern not only reflects the replacement of market dominance but also reveals a deeper structural transformation: by the mid-2040s, the core metabolic system of China's private passenger vehicle fleet—from resource inflows to EOL outflows—will be fully dominated by NEVs.

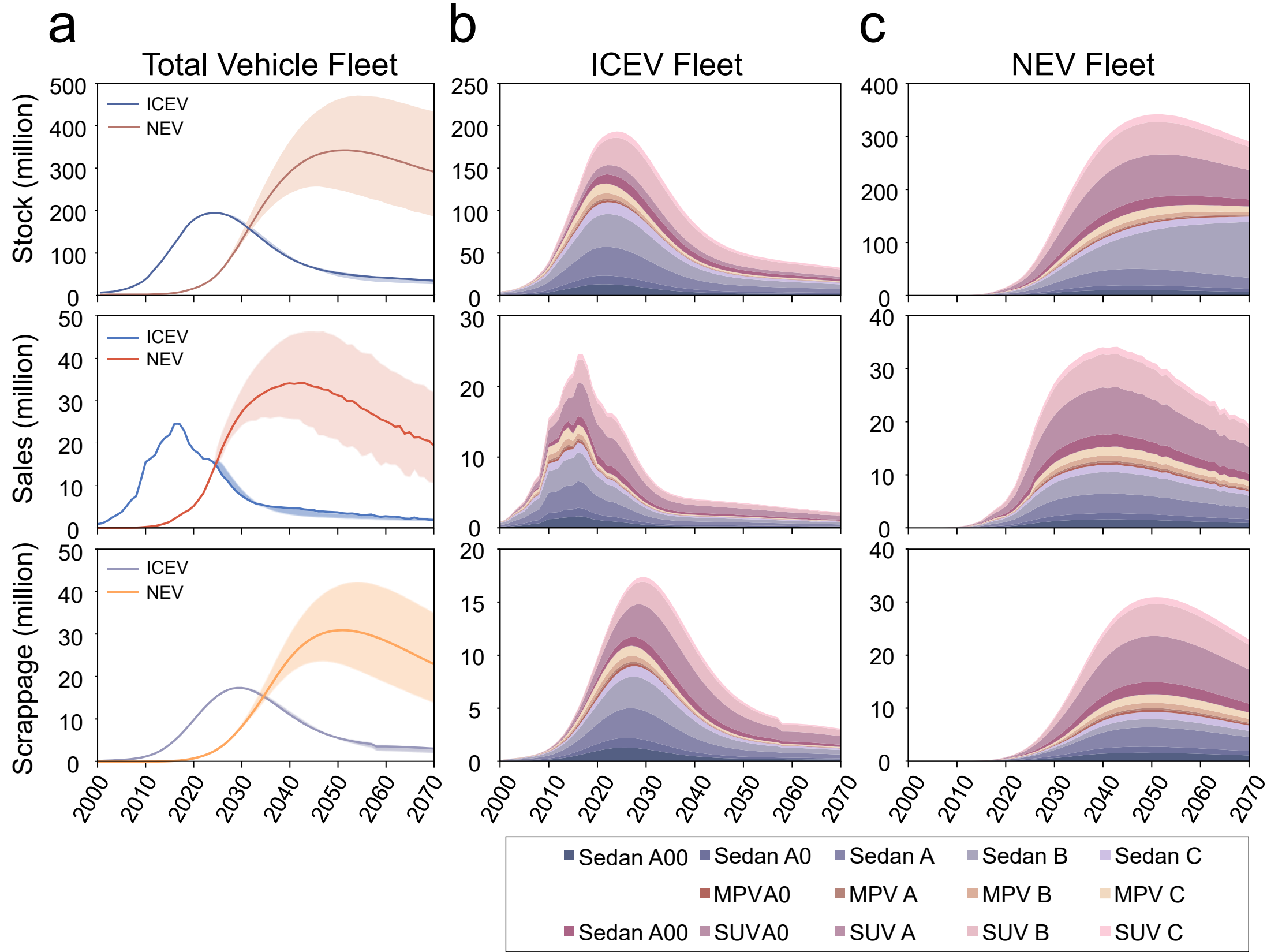

**Fig. 2.** Trends in stock, sales, and scrappage of China's private passenger vehicle fleet, 2000–2070: (a) total fleet under different demand scenarios; (b) ICEV fleet under the BAU scenario; and (c) NEV fleet under the BAU scenario.

Parallel to the transformation of the power system, a trend toward larger vehicles was observed (Fig. 2b–c). The proportion of large vehicles—particularly SUVs—in annual sales was projected to continue increasing. This structural shift further intensifies the material intensity of individual vehicles, thereby increasing demand pressure on metals at the fleet level. Specifically, in the ICEV market, the consumption structure exhibits pronounced upgrading and centralization trends. As the market base evolves, the share of sedans continues to shift away from A00-class microvehicles—which are rapidly being replaced by NEVs—toward A- and B-class models. Moreover, the SUV segment continues to expand, emerging as the primary source of growth in the ICEV market. Within this segment, B-class SUVs account for more than 50% of total SUV sales, constituting the core segment with the highest value and the most intense competition.

In the NEV market, B-class sedans constitute a major share of sedan sales, while in the SUV segment, A- and B-class models together account for approximately 80% of sales, occupying a dominant position. Notably, the share of MPV models in both powertrain markets has remained below 10% over an extended period. This pattern confirms that functional vehicle segmentation is the primary determinant of market scale, whereas differences in powertrain technologies play a secondary role.

In summary, the projected results provide insights into the future dynamics of China's private passenger vehicle fleet, highlighting growth trajectories, market transitions, and the evolution of scrappage volumes. These findings underscore the critical role of NEVs in shaping future fleet size and scrappage patterns, as well as the transition in market dominance from ICEVs to NEVs. Together, these results provide a preliminary response to Question 1 outlined in Section 1.

### *4.2. Evolution of metal demand and recycling potential in the vehicle fleet*

Fig. 3a illustrates the impacts of different scenarios on total metal demand and recycled metal supply. In this study, recycled metals are defined as the fraction of metals recovered from scrapped vehicles and post-consumer waste that re-enter the production system through recycling and reuse processes, constituting the supply of secondary metals. The results show that total metal demand is more sensitive to changes on the demand side, whereas secondary metal supply responds more strongly to changes on the technology side.

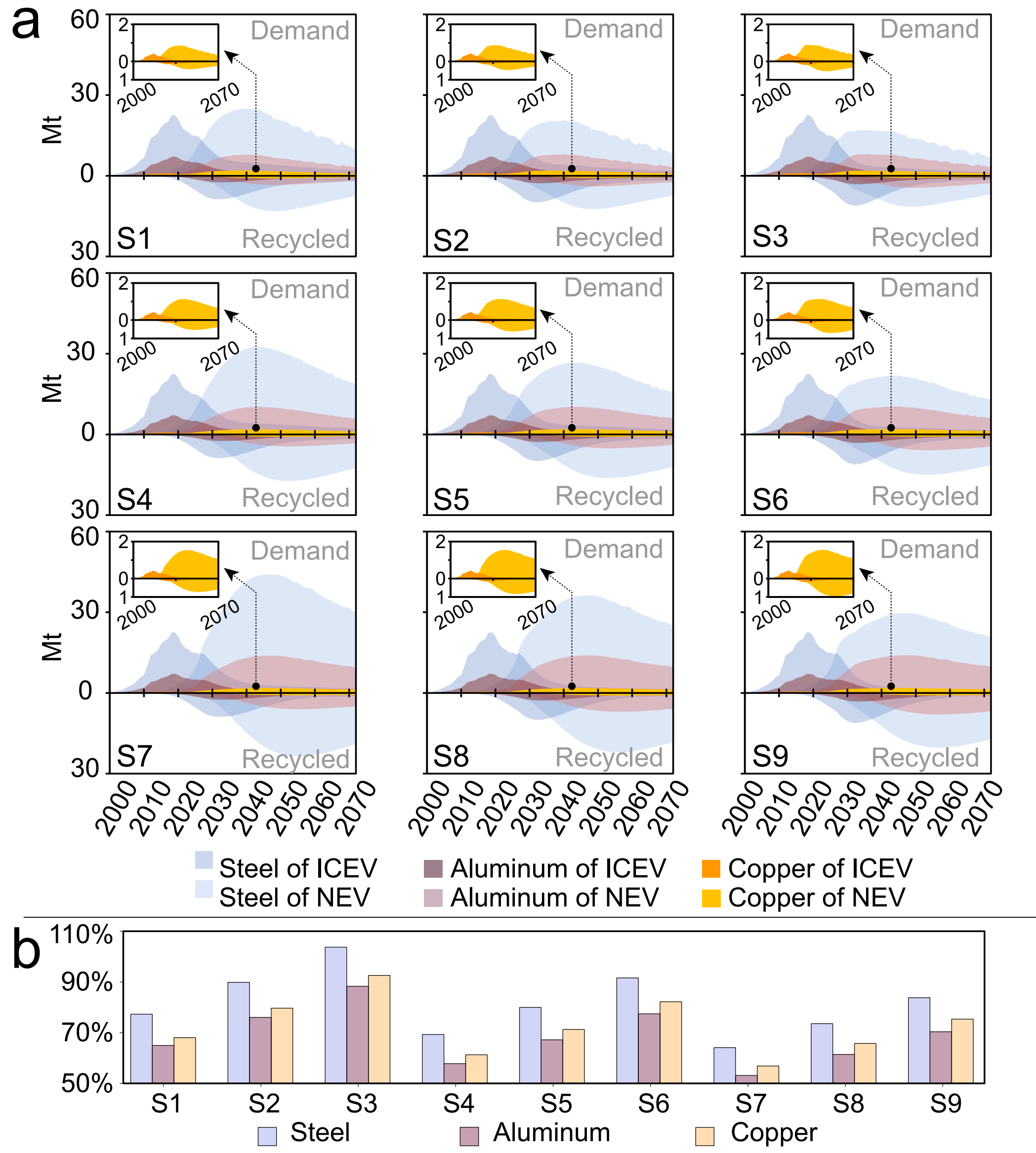

**Fig. 3.** (a) Cumulative metal demand and secondary supply potential under different scenarios during 2000–2070; (b) EOL recycling input rate (EOL-RIR; hereafter referred to as RIR) of steel, aluminum, and copper in 2070.

Scenario analysis reveals that future in-use metal stocks are strongly shaped by demand levels, with significant implications for urban mining and resource management. When technology-oriented parameters are held constant, variations in demand exert a pronounced influence. Relative to the baseline level (2382 megatonnes (Mt) for total demand and 1068 Mt for recycled supply), cumulative outcomes over 2025–2070 change as follows: under high-demand conditions, total metal demand increases by 25.5% and recycled metal supply increases by 23.6%; conversely, under low-demand conditions, total demand decreases by 19.6% and recycled supply decreases by 17.7%. These results indicate that future urban mines are not passive or predetermined outcomes but rather strategic systems that can be actively shaped through demand-side management.

Simulation results for future societal inventories of steel, aluminum, and copper reveal substantial differences across demand levels, highlighting both macro-level drivers and priorities for differentiated resource management. Relative to baseline in-use stocks (309 Mt for steel, 99.3 Mt for aluminum, and 7.9 Mt for copper), projected stocks over 2025–2070 change as follows: under high-demand conditions, in-use stocks increase by approximately 9.5% for steel, 9.2% for aluminum, and 11.4% for copper; under low-demand conditions, they decrease by approximately 4.2%, 3.9%, and 5.1%, respectively. Cross-metal comparisons further underscore distinct strategic priorities. Steel exhibits the largest absolute stock change, reflecting its dominant share of total metal mass and indicating that it should be the primary focus of efforts to improve system-level metal use efficiency. Aluminum shows substantial relative growth, underscoring its rising criticality and the need for robust recycling systems to manage its expanding stock. Although copper experiences the smallest absolute change, its high economic value and carbon intensity per unit mass imply that even modest stock variations have strategic implications for supply security and decarbonization.

Results further indicate that technology-oriented transitions primarily affect the structure of metal supply and the efficiency of metal utilization. Under the baseline demand level, increasing technological ambition from low to high substantially raises recycling input rates (RIRs) for all three metals by more than 20 percentage points by 2070: steel increases from 69.2% to 91.5%, aluminum from 57.7% to 77.4%, and copper from 61.2% to 82.1%. This demonstrates that investment in recycling technologies represents a high-leverage strategy, unlocking urban mining potential by enhancing both the supply capacity and utilization efficiency of secondary metals, while structurally offsetting primary metal demand and reducing long-term reliance on virgin mineral extraction.

Across all technological levels, RIRs exhibit a consistent hierarchy, with steel maintaining higher values than copper and aluminum. This persistent disparity suggests that aluminum resource vulnerability is structural rather than incidental. Its root cause lies in long-term electrification and lightweighting trends in the transport sector, which generate sustained growth in aluminum demand that outpaces advances in recycling technologies and the development of alternative materials.

By 2070, under the high-technology configuration, the RIR of steel reaches 103.6% (119% for ICEVs and 101% for NEVs), the RIR of aluminum reaches 88.2% (107% for ICEVs and 84% for NEVs), and the RIR of copper reaches 92.5% (106% for ICEVs and 91% for NEVs). The steel RIR exceeding 100% indicates that secondary steel supply is sufficient to fully meet the steel demand of private passenger vehicle manufacturing, while also generating surplus secondary steel available for cross-sectoral use. In contrast, although aluminum and copper do not achieve full circular closure, their dependence on primary metal inputs is expected to decline substantially. At the economic and environmental level, nearly 90% of aluminum and copper demand that would otherwise be met by primary production is offset, implying proportional reductions in ore mining, long-distance transportation, and primary smelting. This transition significantly lowers upstream resource extraction costs, energy consumption, and associated carbon emissions. Strategically, these findings confirm that the circular economy is not merely a reactive response to resource constraints, but rather an active pathway for advancing industrial systems toward green, low-carbon, and resource-efficient development. This transition marks a structural shift in the metal supply system from a linear model reliant on "geological mines" to a circular model increasingly supported by "urban mines".

Simulation results further reveal that by 2070, metal RIRs for ICEVs are generally higher than those for NEVs. From the perspective of the DMFA framework, this pattern arises from a structural mismatch between in-use metal stocks and scrappage flows. Although the future passenger vehicle fleet is dominated by NEVs, secondary metal supply remains largely derived from historically accumulated ICEV stocks, causing recycling supply to lag behind demand transitions. Moreover, to meet requirements for lightweighting and energy efficiency, NEVs often exhibit generational differences in metal intensity and metal purity—particularly for aluminum and copper—relative to ICEVs, which further increases technical barriers to secondary metal utilization across powertrain cycles.

A comparison of the marginal benefits of technological measures across demand levels indicates that higher recycling performance ceilings can be achieved under low-demand conditions. Across all simulated scenarios, the highest metal RIR was observed in the S3 scenario, characterized by a combination of low demand and high technological ambition. Under low-demand conditions, although the absolute scale of recyclable metal flows is limited in the short term, high-intensity recycling technologies enable the system to achieve relatively high RIRs, gradually transforming urban mines into stable and high-quality sources of secondary metal supply. This result highlights a critical structural mechanism: a smaller and more stable in-use inventory creates favorable conditions under which efficient recycling technologies can more readily approach closed-loop circulation. In contrast, under high-demand conditions, rapid expansion of in-use stocks dilutes the marginal benefits of technological progress, making it difficult to fully realize gains in resource efficiency.

These findings carry important implications for resource management strategies. Scale moderation followed by efficiency enhancement constitutes the core logic of effective resource optimization. Reliance on EOL recycling measures alone (e.g., the S9 scenario) is insufficient to alleviate resource pressures induced by rapid demand growth, while demand control without concurrent development of circular technologies (e.g., the S1 scenario) fails to establish an efficient recycling system. Only through coordinated demand moderation and technological upgrading can material security be systematically improved.

In summary, the results provide a comprehensive assessment of future metal demand and recycling potential in China's private passenger vehicle fleet, demonstrating how demand pathways and technological advancement jointly shape secondary metal supply. These findings deepen understanding of automotive metal flow dynamics and establish a foundation for future resource management and circular economy strategies aimed at optimizing metal use efficiency. Collectively, these outcomes provide an initial response to Question 2 outlined in Section 1.

*4.3. Embodied carbon emission pathways under alternative scenarios*

To identify the key drivers shaping future carbon emissions within the system, Fig. 4 illustrates the projected carbon emission trajectories across different demand and technology levels.

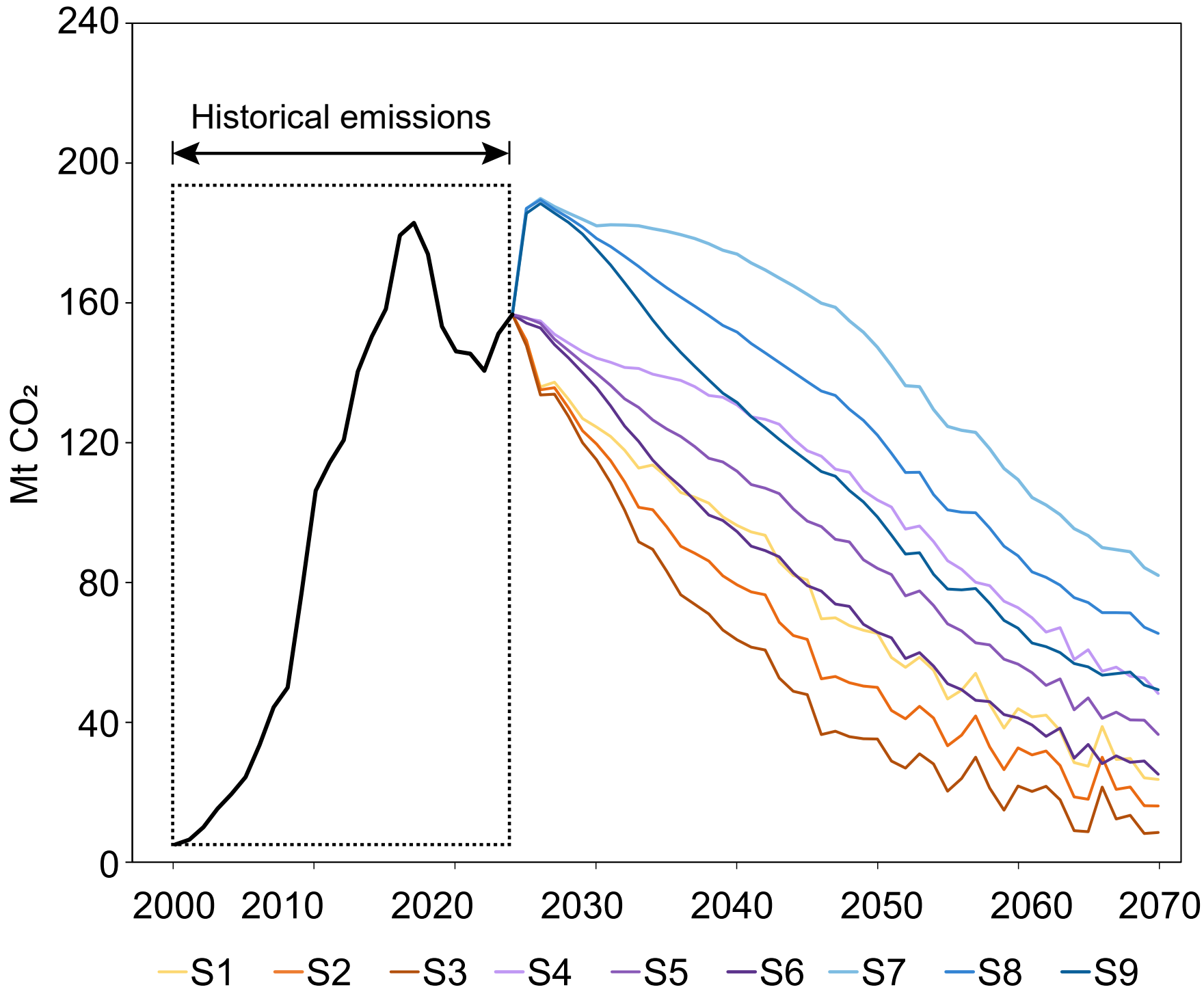

**Fig. 4.** Changes in embodied carbon emissions from vehicle metals in China's private passenger vehicle fleet under nine scenarios, 2020–2070.

Based on a comprehensive analysis of the simulation results, three main conclusions can be drawn. First, technological upgrading is associated with substantial reductions in carbon emissions. When moving from low- to high-technology levels, cumulative emission reductions at different demand levels are as follows: under low demand, emissions decrease by 1051 Mt $CO_2$; under baseline demand, by 1293 Mt $CO_2$; and under high demand, by 1619 Mt $CO_2$. Comparison across demand levels indicates that the magnitude of emission reductions attributable to technological upgrading increases monotonically with demand. Moreover, for any given demand level, higher technology levels consistently result in lower cumulative carbon emissions.

Second, uncontrolled demand growth has the potential to offset the emission reduction benefits of technological progress. Simulation results reveal that demand expansion not only substantially increases baseline production-related emissions but also weakens, or even counteracts, the mitigation effects of technological innovation. For instance, when technology remains at a low level, cumulative carbon emissions increase by 53.4% as demand rises from low to high, from 6009 to 9218 Mt $CO_2$. More importantly, cross-scenario comparisons show that cumulative emissions under the high-demand, high-technology pathway are only approximately 2.6% higher than those under the baseline-demand, low-technology pathway, while emissions under the baseline-demand, high-technology pathway are only about 1.7% higher than those under the low-demand, low-technology pathway. These results suggest that even with advanced technologies, rapid and uncontrolled demand growth can substantially erode the emission reduction benefits of technological upgrading. Consequently, both the pace and scale of demand growth emerge as critical variables in carbon emission control.

Third, the optimal low-carbon pathway depends strongly on the synergy between demand-side management and supply-side technological innovation. The simulation results indicate that the low-demand, high-technology combination achieves the lowest cumulative carbon emissions among all scenarios, at 4958 Mt $CO_2$—representing a 26.4% reduction relative to the baseline-demand, baseline-technology scenario. Compared with the high-demand, low-technology scenario, which exhibits the highest emissions, the corresponding reduction reaches 46.2%. These findings further demonstrate the inherent limitations of relying on a single mitigation approach. Exclusive emphasis on technological research and development risks allowing demand-driven emission growth to undermine mitigation gains, while demand suppression alone may constrain socioeconomic development and eventually encounter diminishing returns.

In summary, the results underscore the critical importance of jointly considering technological upgrading and demand-side management in shaping future carbon emissions from China's private passenger vehicle fleet. Technological progress consistently drives emission reductions, particularly at higher innovation levels; however, unchecked demand growth can significantly weaken these gains. Overall, Section 4.3 provides a direct response to Question 3 outlined in Section 1, highlighting that the most effective low-carbon pathway requires a careful balance between technological advancement and demand control.

## 5. Discussion

### *5.1. Carbon mitigation of individual demand- and technology-oriented measures*

On the basis of a comprehensive scenario analysis, this section examines the independent impacts of individual measures on annual metal production and processing emissions, with all other measures held at their baseline levels. The results are presented in Fig. 5.

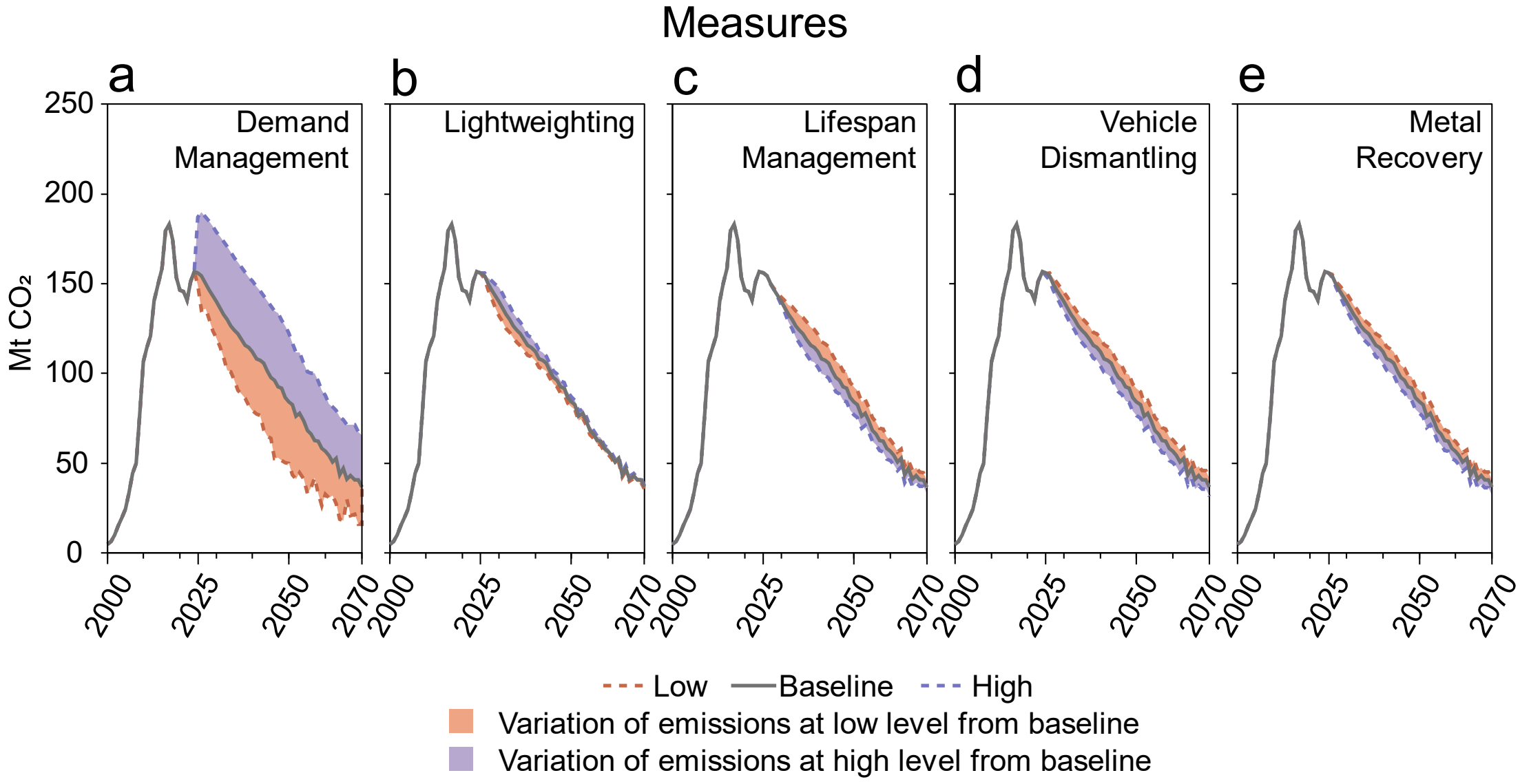

**Fig. 5.** Carbon emissions associated with five individual mitigation measures under low, baseline, and high levels: (a) demand management; (b) lightweighting; (c) lifespan management; (d) vehicle dismantling; and (e) metal recovery.

- **Demand management:** Passenger vehicle ownership demand is identified as a critical driving factor that significantly affects both fleet size and carbon emissions. Owing to China's large population base, differences in ownership levels are rapidly amplified, resulting in substantial macro-level carbon emissions. Peak fleet sizes differ markedly across the three scenarios, reaching 364 million, 420 million, and 530 million vehicles, respectively. By 2070, the number of private passenger vehicles will reach 232 million, 339 million, and 471 million under low-, baseline-, and high-demand levels, corresponding to average annual metal production and processing emissions of 162, 367, and 656 Mt $CO_2$, respectively. These results clearly indicate that shifting from a low-demand to a high-demand pathway not only nearly doubles fleet size by 2070 but also increases average annual carbon emissions by more than fourfold, highlighting the fundamental leverage of demand management as an emission reduction measure in the transport sector.

- **Lightweighting:** Simulation results indicate that the impact of lightweighting on cumulative carbon emissions is relatively limited, with average changes of only about 2.4% compared with the baseline scenario. Assuming all other conditions remain constant through 2070, cumulative steel demand was projected to decrease by 122.7 Mt, with recycled steel declining by 62.6 Mt. In contrast, aluminum demand was expected to increase by 38.0 Mt, with recycled aluminum rising by 16.5 Mt. Due to the increasing penetration of NEVs and the associated growth in electrical components, copper demand was projected to rise by 4.0 Mt, with recycled copper increasing by 1.8 Mt. Compared with other individual measures, lightweighting exhibits relatively low sensitivity in reducing system-level emissions, primarily due to combined constraints related to material mass and energy structure. First, fleet material composition is the dominant factor: under the baseline scenario, steel inventory was projected to reach 193 Mt by 2070, approximately 2.7 times the aluminum inventory of 72.3 Mt. Although aluminum offers higher per-unit emission reduction potential—each tonne of recycled aluminum saves approximately 13 t $CO_2$ relative to recycled steel—its overall impact is constrained by its smaller mass base. Consequently, steel remains the dominant contributor to carbon flows within the vehicle fleet. Second, the emission reduction effectiveness of lightweighting depends strongly on its interaction with power grid decarbonization and the efficiency of aluminum recycling systems; high grid carbon intensity or inefficient recycling pathways can substantially delay or weaken its net benefits.
- **Lifespan management:** The average service life of vehicles is a key indirect factor regulating fleet turnover. Shortening vehicle lifespan accelerates turnover, increasing demand for new vehicle production while also altering the timing of scrappage and recycling flows. Simulation results indicate that by 2070, a one-year reduction in average vehicle lifespan will increase cumulative steel demand and recycling by 5.7% and 7.3%, aluminum demand and recycling by 6.1% and 7.6%, and copper demand and recycling by 6.7% and 8.4%, respectively. This accelerated material cycling intensifies resource use and energy processing throughout the life cycle, ultimately increasing cumulative carbon emissions. Specifically, a one-year reduction in average lifespan results in an increase of approximately 295 Mt $CO_2$ by 2070, whereas a one-year extension reduces cumulative emissions by approximately 252 Mt $CO_2$.

- **Vehicle dismantling and metal recovery:** Simulation results indicate that vehicle dismantling rates and metal recovery rates exhibit similar sensitivities with respect to system-level emission reductions, characterized by an approximately linear relationship. This outcome reflects the model structure, in which these parameters directly govern secondary metal supply. Specifically, a simultaneous 5% increase or decrease in dismantling rates results in an average annual emission reduction change of 6.1%, while an equivalent adjustment in metal recovery rates alters emission reductions by 7.1% relative to the baseline scenario. Holding other parameters constant through 2070, cumulative emission reductions will reach 393 Mt $CO_2$ from improved vehicle dismantling and 327 Mt $CO_2$ from enhanced metal recovery. These findings highlight the systematic influence of dismantling and recycling measures on long-term carbon mitigation. The observed linear sensitivity suggests that improvements in dismantling and recycling efficiency yield near-proportional emission reduction benefits, providing clear quantitative support for setting ambitious and actionable policy targets.

### *5.2. Synergistic carbon mitigation pathways and measure contributions*

The scenario simulation results indicate that from 2000 to 2070, cumulative carbon emission reductions under different development pathways exhibit a continuous upward trend, with substantial differences in both growth rates and final levels. A comparative analysis for 2030 and 2070, selected as representative time points, highlights the evolving divergence among the nine scenarios (see Fig. 6).

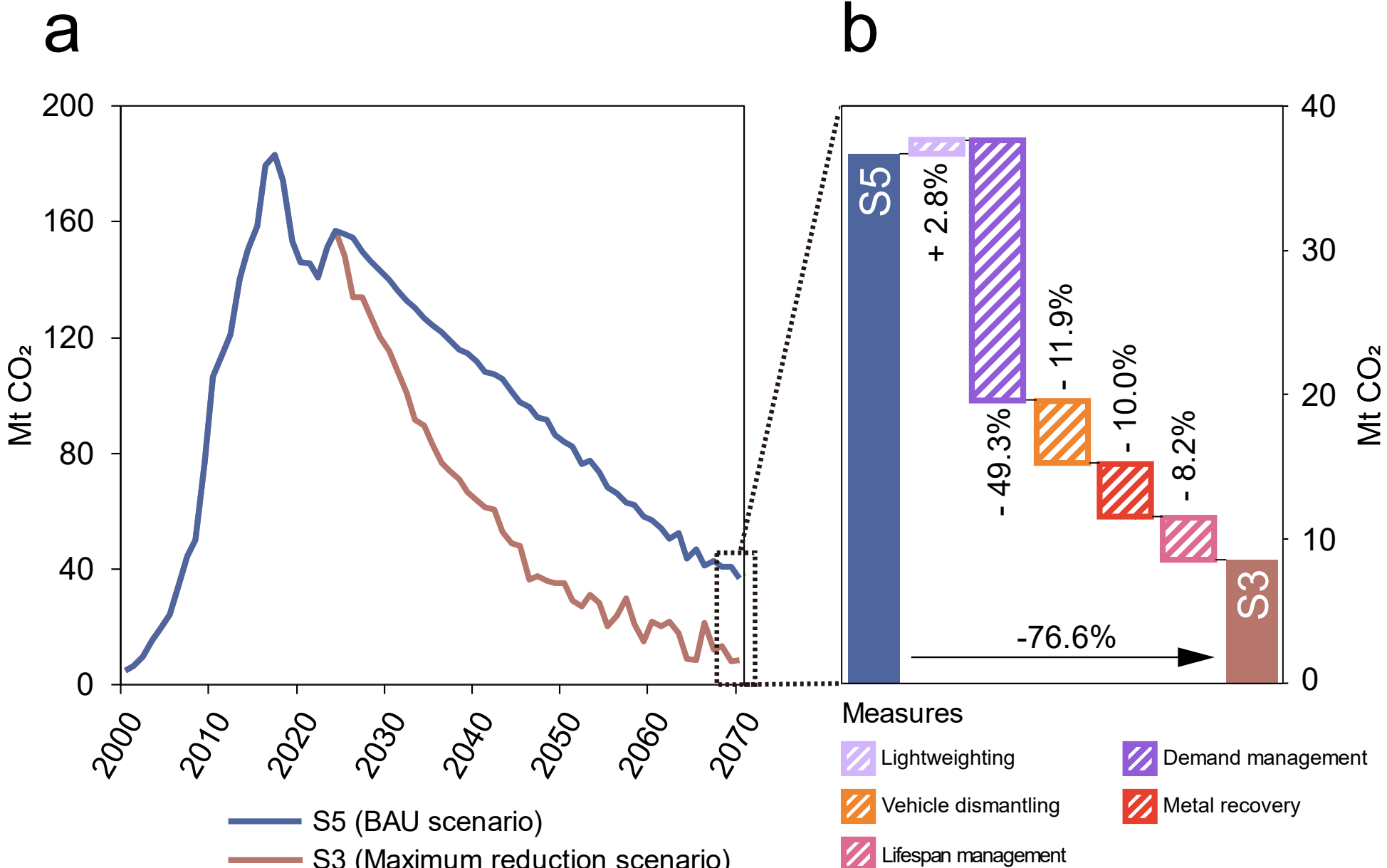

**Fig. 6.** (a) Projected carbon emissions under the BAU scenario and the maximum reduction scenario from 2000 to 2070; (b) contribution shares of five metal use efficiency measures.

In 2030, average annual carbon emissions under the BAU scenario are 140 Mt $CO_2$, whereas emissions under the S3 scenario are 115 Mt $CO_2$, representing a reduction of approximately 17.6%. Over time, the marginal emission reduction benefits of technological progress increase steadily. By 2070, differences among scenarios widen further: average annual emissions under the BAU scenario will decline to 36.7 Mt $CO_2$, while those under the S3 scenario will fall to 8.6 Mt $CO_2$, corresponding to a reduction of approximately 76.6%. The low-carbon transition pathway also exhibits pronounced long-term cumulative effects. By 2070, cumulative emission reductions will reach 1782 Mt $CO_2$, with emissions under the S3 scenario accounting for approximately 26.4% of cumulative emissions under the BAU scenario.

To quantify the contribution of individual measures, measures are ranked in descending order based on their standalone emission reduction potential, which broadly corresponds to their cost-effectiveness from high to low. Synergistic emission reduction effects are then allocated to subsequent measures according to each measure's relative contribution potential. From a dynamic perspective, demand reduction delivers the most significant short-term impact and plays a decisive role in shaping near-term carbon emissions. Over time, the contributions of technology-oriented measures—such as improvements in vehicle dismantling rates, metal recovery rates, and lightweighting—gradually increase, becoming key drivers of the widening emission reduction gaps in the later stages. In addition, high-technology pathways demonstrate stronger incremental emission reduction capabilities in the medium to long term, whereas low-technology pathways exhibit relatively limited mitigation potential, underscoring the importance of technological upgrading for achieving deep decarbonization.

In summary, these findings emphasize that demand management is particularly effective in the early stages of fleet transition, delivering substantial emission reductions by constraining the growth of vehicle ownership. As technological advancements accumulate, they become equally critical in the long term—not only by improving production efficiency but also by enhancing metal recycling and reuse potential, which is essential for sustained emission reduction. Collectively, these results provide a complete response to Question 3 posed in Section 1.

### *5.3. Policy recommendations*

Based on these findings, policymakers are encouraged to treat passenger transport demand management as a strategic pillar alongside the promotion of NEVs and to integrate it into national climate and sustainable development planning, as has already been adopted in several countries [52]. Existing studies have emphasized that restraining the growth of vehicle ownership and overall mobility demand is critical for mitigating carbon emissions in the transport sector [53]. This study further demonstrates that demand management is among the most effective emission reduction measures, as unchecked demand growth can substantially undermine the mitigation benefits of technological improvements [54], even under advanced fleet electrification and vehicle efficiency scenarios. This conclusion is directly supported by the results in Section 5.1 and the decomposition analysis in Section 5.2, which show that demand-side measures deliver the largest and most immediate emission reductions and largely determine the long-term emission trajectory of the vehicle fleet. To translate the modeled demand pathways into actionable policy instruments, demand moderation can be pursued through a portfolio of mutually reinforcing measures, including vehicle ownership controls, pricing-based instruments, and modal shift policies [55]. Specifically, ownership-oriented measures—such as license plate quotas, vehicle registration lotteries, and differentiated registration fees based on vehicle size and energy type—can directly constrain fleet expansion in large metropolitan areas. Pricing reforms, including congestion charging, parking pricing, fuel or electricity taxation linked to carbon intensity, and distance-based road user charges, serve to internalize the external costs of private vehicle use and further discourage excessive mobility demand. In parallel, sustained modal shift can be facilitated through long-term investments in high-capacity public transport systems, alongside improvements in the safety, accessibility, and continuity of non-motorized transport infrastructure, thereby providing viable alternatives to private vehicle ownership and use.

In the context of rapid urbanization, promoting balanced land-use planning and high-capacity public transport systems [56]—such as metro systems, urban rail, and bus rapid transit—combined with safe and accessible non-motorized transport infrastructure, can effectively reduce reliance on private vehicles and facilitate the transition toward low-carbon mobility [57, 58]. Moreover, previous studies indicated that shifting passenger transport from private vehicles to public modes (e.g., metro and rail) and freight transport from trucks to rail and waterways can reduce $CO_2$ emissions by up to 35% relative to the reference scenario by 2060 [59]. Complementary policies that accelerate fleet electrification, improve vehicle efficiency, and promote NEV adoption [60] can further enhance overall emission reduction potential when coordinated with demand-side interventions.

Importantly, the dynamic results of this study indicate that demand management plays a particularly critical role in the near to mid term, whereas technological measures—such as fleet electrification, material efficiency improvements, and recycling system maturation—become increasingly important over the longer term. This temporal shift has direct implications for intergenerational equity: demand-side policies often impose higher short-term behavioral and economic costs on current users, whereas the benefits of technological progress accrue predominantly to future generations. To address this imbalance, early-stage demand management should be accompanied by compensatory mechanisms, such as revenue recycling from congestion charges or carbon pricing to support public transport investment, affordability measures for low-income households, and targeted subsidies for clean mobility options, thereby promoting a socially equitable transition across generations.

Notably, increasing the share of aluminum and copper in vehicle bodies may raise primary demand for these carbon-intensive metals, potentially offsetting part of the expected emission reductions. This challenge is consistent with previous studies highlighting the trade-off between lightweighting and increased demand for high-carbon materials [61], particularly when their production remains carbon intensive [62]. Results from Section 5.1 indicate that the net carbon benefit of lightweighting depends critically on two conditions: the establishment of efficient closed-loop recycling systems for automotive aluminum and the decarbonization of metal production processes. Without effective recycling, the emission benefits of substituting steel with aluminum are significantly diminished. Given the long-term horizon to 2070, recycling and production decarbonization targets should be designed in a staged and adaptive manner. For example, moderate but enforceable dismantling and recovery targets could be implemented in the near term to avoid excessive compliance costs, while progressively stricter targets could be introduced as recycling technologies mature and secondary metal markets expand. Such a phased approach would help distribute costs more evenly across generations while ensuring that long-term mitigation potential is fully realized. Consequently, lightweighting strategies must be integrated with robust recycling infrastructure and material system design to maximize carbon mitigation. As shown by this study, expanding the use of secondary metals through recycling can substantially enhance the long-term carbon reduction potential of lightweighting. Policymakers should therefore prioritize investments in closed-loop recycling systems for automotive metals, incentivize green production technologies [63], and promote cross-sectoral synergies to accelerate technological innovation [64]. Such coordinated measures ensure that lightweighting contributes to emission reductions while minimizing upstream carbon burdens.

Although the emission reduction potential of extending vehicle lifespan is relatively modest compared with demand management and technological upgrading, lifespan extension remains an effective strategy for reducing resource consumption and improving metal use efficiency [65].

As shown in Section 5.2, changes in vehicle lifespan exert a moderate but non-negligible influence on cumulative carbon emissions. Policy interventions aimed at extending vehicle service life can therefore complement broader decarbonization strategies. Product durability can be enhanced through technical standards that emphasize maintainability, modular design, and upgradeability of key components, guiding manufacturers toward long-term design strategies. Financial incentives—such as tax reductions or targeted subsidies [66]—can encourage consumers to maintain vehicles and reduce premature scrappage [67]. Automakers can also be incentivized to improve durability through stricter quality standards and design-for-longevity principles, supported by advances in modular manufacturing and maintenance technologies. In parallel, strengthening regulation and certification of the second-hand vehicle market can help maximize resource utilization, extend vehicle lifespans, and reduce life-cycle carbon emissions.

Finally, based on the results related to vehicle dismantling and metal recovery, three levels of policy recommendations are proposed. First, to realize the recycling potential identified in this study, mandatory dismantling and material recovery targets should be explicitly defined along a temporal pathway (e.g., short-, medium-, and long-term milestones), rather than as static objectives, and integrated into broader automotive decarbonization and green development strategies [68]. This recommendation is directly supported by the analyses in Sections 5.1 and 5.2, which reveal an approximately linear relationship between dismantling and recovery rates and long-term emission reductions, underscoring their strong controllability and scalability. Second, to address material quality and efficiency constraints identified in the flow analysis, ELV recycling networks should be strengthened by promoting formal dismantling through extended producer responsibility schemes and financial incentives, while supporting technological development in component identification, material sorting, and efficient separation of composite metals [69, 70]. These measures can enhance the supply of high-quality secondary metals, reduce dependence on primary resources, and increase circular material use within the automotive sector [71]. Third, to internalize and reward the substantial carbon reduction benefits of recycling, a comprehensive carbon accounting and management framework for automotive metals should be established. Explicitly recognizing recycling-related emission reductions within carbon markets, green finance mechanisms, and performance-based subsidy schemes would enable early investments in recycling infrastructure to generate near-term economic returns, while delivering long-term climate benefits for future generations. Integrating quantified recycling-related emission reductions into national carbon markets and green finance mechanisms would provide direct economic incentives for the use of secondary metals and accelerate the transition toward a circular, low-carbon automotive industry [72, 73].

## 6. Conclusion

This study developed a transferable, data-driven modeling framework to assess metal flows and embodied carbon emissions in China's private passenger vehicle fleet under long-term transition pathways. The framework integrated fleet dynamics, vehicle metal stocks and flows, and embodied carbon emissions, while explicitly accounting for cohort-level characteristics and the interactions among metals, energy use, and emissions. Focusing on steel, aluminum, and copper, the model simulated stock-flow dynamics from 2025 to 2070 and applied multi-scenario analyses to quantify the individual and synergistic impacts of demand-oriented and technology-oriented metal use efficiency measures on carbon reduction. The key findings of this study are summarized below.

### *6.1. Principal findings*

- **China's private passenger vehicle fleet would peak at 327–507 million vehicles around mid-century, accompanied by a pronounced restructuring of fleet composition and inflow-outflow dynamics.** With accelerated electrification, ICEV stocks would decline to 24.5–32.7 million vehicles by 2070, while NEV stocks would expand to 184.6–433.4 million vehicles, progressively replacing ICEVs as the dominant component of both in-use stocks and scrappage. On the inflow side, NEVs have surpassed ICEVs in annual sales in the mid-2020s, and total vehicle sales would subsequently peak at approximately 34.1 million vehicles in the early 2040s. Consequently, total vehicle scrappage would peak around 2046 at approximately 37.0 million vehicles per year, with NEVs accounting for about 84% of EOL vehicles. These trends indicate a system-level transition toward an NEV-dominated fleet metabolism by the mid-2040s. In parallel, vehicle composition continues to shift toward larger models—particularly B-segment SUVs—in both the ICEV and NEV markets.
- **From 2025 to 2070, cumulative metal demand would range from 1914 to 2990 Mt across different demand levels, while cumulative secondary metal supply would reach 879–1320 Mt.** Relative to the baseline demand level, the low-demand scenario would reduce total metal demand by approximately 467 Mt, whereas the high-demand scenario would increase demand by about 607 Mt, indicating that demand intensity primarily governs the scale of metal inflows and outflows. In contrast, technological upgrading mainly reshapes the supply structure. Under the baseline demand level, improved recycling performance would increase metal RIRs by more than 20 percentage points by 2070, enabling secondary steel to fully meet new manufacturing demand and allowing aluminum and copper cycles to approach near closure. These results demonstrate that metal flow magnitude is largely demand-driven, whereas urban mining potential is unlocked through improvements in recycling efficiency.

- **Correspondingly, cumulative embodied carbon emissions from vehicle metals by 2070 would range from 4958 to 9218 Mt $CO_2$, with technological upgrading reducing emissions by 1051–1619 Mt $CO_2$ depending on demand levels.** Although higher demand enlarges the absolute mitigation potential of technological measures, cumulative emissions would still increase by 53.4% when moving from low- to high-demand levels under low-technology conditions, highlighting the dominance of scale effects. Cross-scenario comparisons further show that cumulative emissions under a high-demand, high-technology pathway would exceed those under a baseline-demand, low-technology pathway by only 2.6%, while baseline-demand, high-technology emissions are just 1.7% higher than those under low-demand, low-technology conditions. This near-equivalence across contrasting pathways underscores strong substitution effects between demand moderation and technological progress and demonstrates that unchecked demand growth can substantially erode, or even offset, cumulative mitigation gains from technological upgrading. In the collaborative scenario, demand management would contribute 64.3% of total emission reductions, followed by vehicle dismantling (15.5%), metal recovery (13.1%), lifespan management (10.7%), and lightweighting (3.6%). Demand-side measures exert the strongest short-term effects by constraining fleet expansion, while technology-oriented measures play an increasingly important role over the medium to long term by expanding secondary metal supply and reducing production-related emissions.

### 6.2. Upcoming works

Building on the results and limitations of this study, future research should prioritize refining the current framework to explicitly account for regional heterogeneity in China. First, from a theoretical perspective, subsequent studies could extend the national-scale analysis to capture substantial provincial variations in electricity grid carbon intensity, vehicle ownership patterns, recycling infrastructure, and policy implementation. Such efforts would enhance the academic understanding of spatial heterogeneity in vehicle recycling systems and their associated carbon mitigation potential. Second, from a practical perspective, future work could develop a province-level modeling framework that disaggregates vehicle stocks, end-of-life flows, metal recovery pathways, and emission factors. By incorporating dynamic, region-specific electricity and transportation emission profiles, this approach would enable a more accurate assessment of regional recycling performance and secondary metal supply.

## Appendix

Please find the appendix in the supplementary materials (e-component) attached to the file "************.pdf". It is also available at the following link:

https://www.researchgate.net/publication/403025127_The_latest_version_of_the_appendix_for_the_revised_manuscript_entitled_Decarbonizing_China's_private_passenger_vehicles_A_dynamic_material_flow_assessment_of_metal_demands_and_embodied_emissions

## Acknowledgments

J.L. acknowledges support from the Chongqing Association for Science and Technology Think Tank Research Project (No. 2025KXKT40). K.Y. acknowledges support from the Beijing Natural Science Foundation (Grant No. 9254035). Contribution from N.Z. is supported by the Energy Demand Changes Induced by Technological 30 and the Social Innovations network, an initiative coordinated by the Research Institute of Innovative Technology for the Earth and the International Institute for Applied Systems Analysis and funded by the Ministry of Economy, Trade, and Industry, Japan.

## Declaration of interests

The authors declare that they have no competing interests.